\documentclass{iopart}
\usepackage{iopams}
\usepackage{dsfont}
\usepackage{amsthm}
\usepackage{graphicx}
\usepackage{bbm}
\usepackage{cite}
\usepackage{setspace}

\newtheorem{Th}{Theorem}
\newcommand{\vect}[1]{\mathbf{#1}}

\begin{document}
\title[Signatures of three coalescing eigenfunctions]{Signatures of three coalescing eigenfunctions}

\author[G~Demange \& E~M~Graefe]{Gilles~Demange and Eva-Maria Graefe}

\address{Department of Mathematics, Imperial College London,
London SW7 2AZ, UK}

\begin{abstract}
Parameter dependent non-Hermitian quantum systems typically not only possess eigenvalue degeneracies, but also degeneracies of the corresponding eigenfunctions at exceptional points. While the effect of two coalescing eigenfunctions on cyclic parameter variation is well investigated, little attention has hitherto been paid to the effect of more than two coalescing eigenfunctions. Here a characterisation of behaviours of symmetric Hamiltonians with three coalescing eigenfunctions is presented, using perturbation theory for non-Hermitian operators. Two main types of parameter perturbations need to be distinguished, which lead to characteristic eigenvalue and eigenvector patterns under cyclic variation. A physical system is introduced for which both behaviours might be experimentally accessible.
\end{abstract}
\vspace{0.4cm}

\section{Introduction}
The coalescence of eigenvalues and the corresponding eigenvectors of operators at so-called exceptional points (EPs) in parameter space plays a crucial role in various branches of physics, reaching from classical mechanics \cite{Seyr03,Kiri07} and optics \cite{Panc55,Berr98b,Berr03,Klai08,Wier08,Dett09,Lee09,Long10,Grae11} to quantum mechanics \cite{Mois11book,Berr04}. Since Hermitian operators have a complete set of eigenfunctions, EPs are a phenomenon that can only occur for non-Hermitian operators. In quantum mechanics, however, the Hamiltonian is traditionally demanded to be Hermitian for the description of closed systems. Non-Hermiticity, and with it EPs, can enter quantum mechanics in different ways: Most importantly, as an effective description of open systems  \cite{Mois11book}, or in the form of PT-symmetric quantum mechanics \cite{Bend99a} aiming at a generalised description of closed systems \cite{Bend02b}. Further, the analogy between the Schr\"odinger and the Helmholtz equations allows beautiful experimental realisations of non-Hermitian quantum systems in microwave cavities \cite{Demb01,Demb04,Diet07,Diet11}, and the fact that the spatial propagation of light in wave guide arrays is described by an equation similar to the time-dependent Schr\"odinger equation led to experimental realisations of non-Hermitian quantum dynamics in optical systems \cite{Guo09,Ruet10}.

EPs have been investigated in one form or another for decades \cite{Panc55,Kato66book,Bend69,Mois78}, and they have attracted considerable attention in recent years, with the increasing interest in non-Hermitian quantum theories. The behaviour of eigenvalues and eigenvectors under cyclic parameter variation in the presence of EPs has been investigated in detail theoretically \cite{Heis99,Heis01,03crossing,Mail05,Mehr08,Cart07}, and the predictions have been confirmed experimentally in microwave cavities \cite{Demb01,Demb04,Diet11}. The influence of EPs on dynamics \cite{Panc55,Heis10,Zhen10,Grae11,Cart11}, leading to typical quadratic behaviour for two coalescing eigenfunctions could also be demonstrated in experiments \cite{Diet07}. The experimental studies on EPs were so far confined to systems which are governed by wave equations that allow to mimic the quantum behaviour, such as the above mentioned microwave cavities, laser cavities, or optical wave guide structures. Only recently there have been suggestions for possible experimental observations of EPs in genuine quantum systems \cite{Cart07,Cart11,Else11,Atab11}. 

While EPs at which two eigenfunctions coalesce are well investigated, little attention has hitherto been paid to the coalescence of more than two eigenfunctions. Among the few exceptions are \cite{Andr07,Cart09,Heis08,08PT,Heis10}. The typical effect on the dynamics at an EP of more coalescing eigenfunctions is a straight forward generalisation of the behaviour at a two fold degeneracy \cite{Heis10}, the influence on the cyclic behaviour, on the contrary, is more elaborate. In \cite{Heis08} an interesting chiral behaviour of the eigenfunctions in the neighbourhood of three coalescing eigenfunctions has been reported. The topological structure of three interacting eigenfunctions in the form of neighbouring EPs of each two eigenfunctions has been investigated in \cite{Ryu11}. In this context it was found that the interchange of the eigenfunctions when encircling several EPs can be quite different from the expected behaviour around a degeneracy of three eigenfunctions, which is often identified with a third root branch point in the literature \cite{Heis08,Cart09}. However, while the third root behaviour is \textit{typical}, in the sense specified below, it is by no means the only possible structure \cite{Kato66book,Moro97,Ma98,08PT}. Thus, although the coalescence of more than two eigenfunctions  is less commonly found in physical systems than the standard EP due to the higher codimension \cite{Heis08}, the nontrivial features associated with such degeneracies make it desireable to gain a deeper understanding into the related structures. The purpose of the present paper is to initiate these investigations.

Here we present a study of the quantum signatures in the neighbourhood of three coalescing eigenvalues for symmetric Hamiltonians. We focus in particular on the behaviour of eigenvalues and eigenvectors under parameter variation, and the resulting interchanges between eigenfunctions and possible additional geometric phases. Using generalised perturbation theory for non-Hermitian operators we show that there are two distinct fundamental scenarios. We suggest an experimental setup, consisting of three coupled wave guides, in which both behaviours could in principle be observable. 

The paper is organised as follows: We begin with a short overview on the mathematical definition of exceptional points and the related Jordan block theory in Section \ref{sec_EP_Jordan}. In Section \ref{sec_pert} we review the general series expansion of eigenvalues and eigenvectors in dependence on a complex perturbation parameter. We analyse the resulting patterns of eigenvalues under variation of parameters, in particular, for the cyclic variation of a complex parameter. We then turn to the behaviour of the eigenvectors under cyclic parameter variation, and the corresponding geometric phases in Section \ref{sec_geo}. We present a physical example in Section \ref{sec_ex}. In this context we draw attention to possible problems when relying on adiabatic assumptions in experiments \cite{Uzdi11,Berr11,Berr11b}. In Section \ref{sec_gen} we briefly comment on a more general two-parameter perturbation. We summarise our results in Section \ref{sec_dis}. 

\section{Exceptional Points and Jordan block theory}
\label{sec_EP_Jordan}
Let us begin with a brief summary of the mathematics of exceptional points, where we also introduce the notations used here. For details on Jordan chains and related topics see, e.g., \cite{Seyr03}. For convenience we work in a finite-dimensional Hilbert space. Hamiltonians acting on this space are represented by complex non-Hermitian matrices depending on a number of system parameters. An exceptional point (EP) is a point in parameter space at which at least two eigenvalues and the corresponding eigenvectors of the Hamiltonian coalesce. That is, at these points the Hamiltonian is not diagonalisable, but it can be brought to Jordan normal form via a similarity transformation $\hat H=R\hat JR^{-1}$, where $\hat J$ is block diagonal, consisting of Jordan blocks $\hat J_n$ associated with the eigenvalues $\lambda_n$:
\begin{equation}
\hat J_n=\left(\begin{array}{ccccc}
   \lambda_n & 1 & 0 & \cdots &0 \\
   0 & \lambda_n & 1 & \cdots & 0\\
   \vdots & \vdots & \vdots& \ddots & \vdots \\
   0 & 0 & 0 & \cdots & 1 \\
   0 & 0 & 0 & \cdots & \lambda_n
 \end{array}\right).
\end{equation}
If $N$ eigenvalues and their corresponding eigenvectors coalesce while the remaining eigenvalues are nondegenerate, we refer to the configuration as an EP$N$. At an EP$N$ the Jordan normal form of $\hat H$ contains one Jordan block of length $N$ and other than that only blocks of size one, corresponding to the nondegenerate eigenvectors.

If an eigenvalue is $N$-fold degenerate with only one linearly independent eigenvector, there exist $N-1$ associated eigenvectors that form the so-called Jordan chain:
\begin{equation}
(\hat H-\lambda)|u_0\rangle=0,\quad (\hat H-\lambda)|u_j\rangle=|u_{j-1}\rangle ,
\end{equation}
where $(\hat H-\lambda)|x\rangle=|u_{N-1}\rangle$ is insoluble. 

When dealing with non-Hermitian Hamiltonians one further has to distinguish left and right eigenvectors, where the left eigenvectors are commonly defined as the right eigenvectors of the adjoint operator corresponding to the complex conjugate eigenvalue. For a nondegenerate spectrum the left and right eigenvectors form a biorthogonal basis. In the degenerate case in addition to the eigenvectors the left and right Jordan chains are necessary to provide a basis for the state space. In analogy with the left eigenvector the left Jordan chain is defined as the Jordan chain of the adjoint operator $\hat H^{\dagger}$:
\begin{equation}
(\hat H^{\dagger}-\bar\lambda)|v_0\rangle=0,\quad (\hat H^{\dagger}-\bar\lambda)|v_j\rangle=|v_{j-1}\rangle,
\end{equation}
where $(\hat H^{\dagger}-\bar\lambda)|x\rangle=|v_{N-1}\rangle$ is insoluble.  

In the present study we confine ourselves to the investigation of symmetric matrices as a first step towards the understanding of the physics of three coalescing eigenvectors. Note, however, that the generalisation to nonsymmetric matrices can lead to qualitatively new effects, in particular, with respect to the geometric phase, as has been investigated in detail for EP2s \cite{Mehr08,Diet11}. In the case of symmetric matrices the left eigenvectors and Jordan chain vectors reduce to the complex conjugates of the right eigenvectors and Jordan chain: $|v_j\rangle=|\bar u_j\rangle$. Thus the scalar product between left and right eigenvectors reduces to the standard inner product of real Hilbert spaces, $\langle v_j|u_k\rangle=u_j^T u_k$, which in the present context is often referred to as the $c$-product \cite{Mois11book}.  For convenience we will from now on omit the bracket notation and denote the right eigenvectors and Jordan chains simply by $\vect{u_j}$. At an EP it holds that $\vect{u_0}^T \vect{u_0}=0$, which can be interpreted as \textit{self-orthogonality} \cite{Mois11book}.

The Jordan chain vectors are \textit{a priory} not uniquely defined, however, with additional normalisation conditions the ambiguity can be removed (up to a sign) \cite{Seyr03}. In addition to the self-orthogonality condition they fulfil a number of further relations, depending on their length. 
In what follows we shall mainly be concerned with EP2s and EP3s, thus we will confine ourselves to the discussion of Jordan chains of length two and three here. 
For a Jordan chain of length two we have the self-orthogonality condition
\begin{equation}
\label{eqn_J1_EP2}
\rm{EP2}:\quad \vect{u_0}^T\vect{u_0}=0.
\end{equation} 
Further, it is convenient to impose the conditions
\begin{equation}
\label{eqn_J2_EP2}
\rm{EP2}:\quad \vect{u_0}^T\vect{u_1}=1 \quad \rm{and}\quad \vect{u_1}^T\vect{u_1}=0,
\end{equation}
which uniquely determine the Jordan chain (up to a sign). 
For an EP3, that is, a Jordan chain of length three, these relations change and one finds instead
\begin{eqnarray}
\label{eqn_J1_EP3}
\rm{EP3}:\quad \vect{u_0}^T\vect{u_0}=0,\quad  \vect{u_0}^T\vect{u_1}=0,\quad \rm{and} \quad \vect{u_0}^T\vect{u_2}= \vect{u_1}^T\vect{u_1}.
\end{eqnarray}
To remove the ambiguity the following relations are imposed:
\begin{equation}
\label{eqn_J2_EP3}
\rm{EP3}:\quad \vect{u_0}^T\vect{u_2}=1, \quad \rm{and}\quad \vect{u_2}^T\vect{u_1}=0=\vect{u_2}^T\vect{u_2}.
\end{equation}

As we shall see below the structure of the Jordan chain \textit{at} the EP is the key for the topological phases related to encircling an EP.  The typical interchange of eigenvalues and eigenfunctions, which is often discussed as part of the same phenomenon, is in fact due to the perturbative behaviour \textit{around} the EP, which we shall discuss in the following section.
\begin{figure}[htb!]
\centering%
\includegraphics[width=6cm]{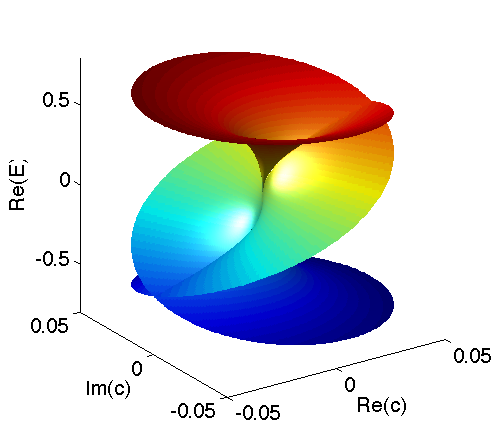}
\includegraphics[width=6cm]{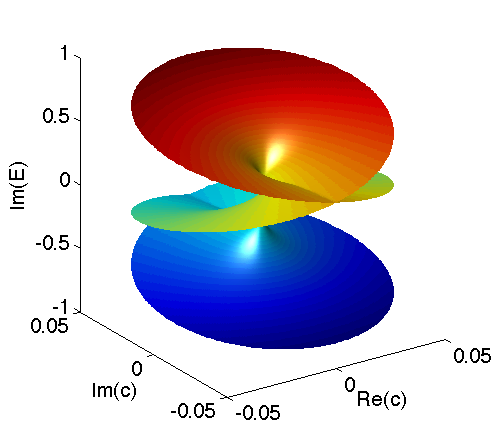}
\includegraphics[width=6cm]{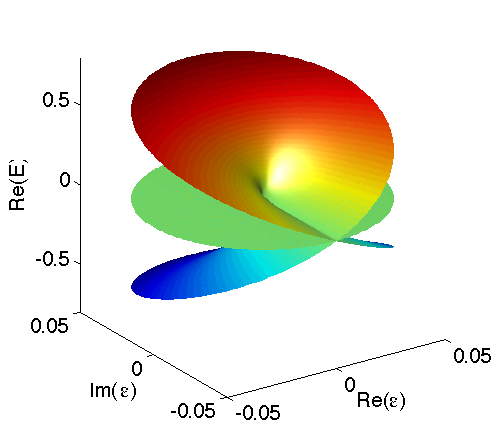}
\includegraphics[width=6cm]{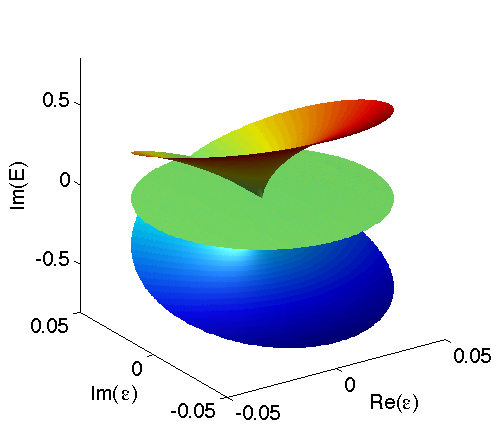}
\caption{Typical cases for the two possible perturbation scenarios of the eigenvalues around an EP3. Shown here are the real (left column) and the imaginary (right column) parts of the eigenvalues of a $3\times 3$ matix as a function of a complex perturbation parameter in the neighbourhood of an EP3. The upper panel shows the familiar third-root behaviour. The lower panels show an example for which a square-root perturbation is observed for two of the eigenvalues, whereas the third eigenvalue is constant. This behaviour is usually connected to EP2s. The examples depicted here correpond to the Hamiltonian (\ref{eqn_example_Ham}) with $\gamma=1=v$ and $a=b=2c$ (top), and $a=-b=2\varepsilon$ (bottom).}
\label{fig1}
\end{figure}

\section{Perturbation of parameters around exceptional points}
\label{sec_pert}
One of the most characteristic features of EPs is the behaviour of the eigenvalues under a perturbation in the neighbourhood of the EP. For Hermitian operators the perturbed eigenvalues can be expanded into a Taylor series in the perturbation parameter. This behaviour holds also for non-Hermitian operators, as long as the perturbation is performed around a nondegenerate eigenvalue or a degenerate eigenvalue corresponding to nondegenerate eigenvectors \cite{Kato66book}. If, however, a system is perturbed around an EP the series expansion is in general a Puiseux series that might involve fractional exponents. Note, however, that contrary to common folklore, this does not necessarily have to be the case. Instead, the following theorem on the ``typical behaviour'' of eigenvalues under a perturbation around an EP$N$ holds \cite{Ma98}:

\begin{Th}
\label{Th1}
Let $\hat A$ be a linear perturbation of $\hat A_0$:
\begin{equation}
\hat A=\hat A_0+z \hat A_1,
\end{equation}
where $z\in\mathds{C}$, and $\hat A_0$ is a full Jordan block of size $N$ with the
eigenvalue $\lambda_0$. If the perturbation matrix $\hat
A_1=(a_{kj})$ (the matrix representation of $\hat A_1$ in the Jordan
basis of $\hat A_0$) is such that $a_{n1}\neq 0$, then the set of
perturbed eigenvalues of $\hat A$ consists of the $N$ branches of an
analytic function, i.e.
\begin{equation}
\lambda_{\sigma}(z)=\lambda_0+\sum_{\mu=1}^{\infty}
\lambda_{\mu \sigma}z^{\mu/N},
\end{equation}
where $\lambda_{1\sigma}=a_{n1}^{1/N}$, and where $\sigma$ labels the
different branches of the $N$-th root.
\end{Th}
The condition $a_{n1}\neq 0$ is crucial here. Although it is often argued that ``most perturbations fulfil this condition'', it turns out that this is not true for many physical situations, as for example in lattice models described by band matrices.
If, however, this condition is not fulfilled, the perturbed eigenvalues can split into different cycles each of which consists of the branches of an analytic function of the form
\begin{equation}
\lambda_{\sigma}(z)=\lambda_0+\sum_{\mu=1}^{\infty}
\lambda_{\mu \sigma}z^{\mu/k},
\end{equation}
where $k<N$, and the different values of $k$ sum up to $N$ \cite{Kato66book,Moro97}.
In the case of symmetric matrices, a similar expansion is valid for the eigenvectors.
\begin{figure}[tb!]
\centering%
\includegraphics[width=4cm]{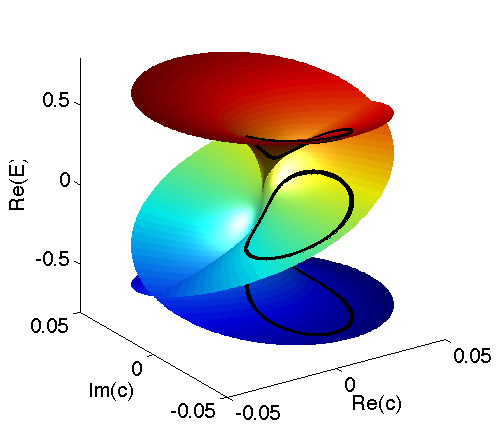}
\includegraphics[width=4cm]{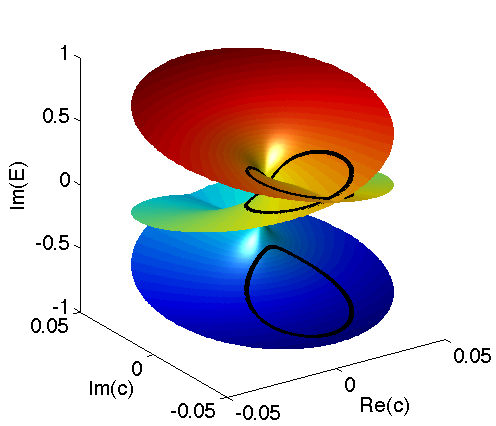}
\includegraphics[width=4cm]{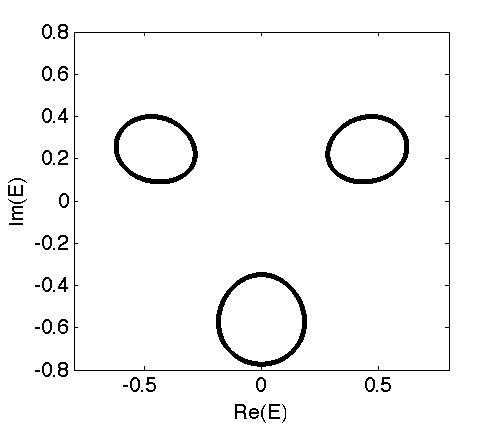}
\includegraphics[width=4cm]{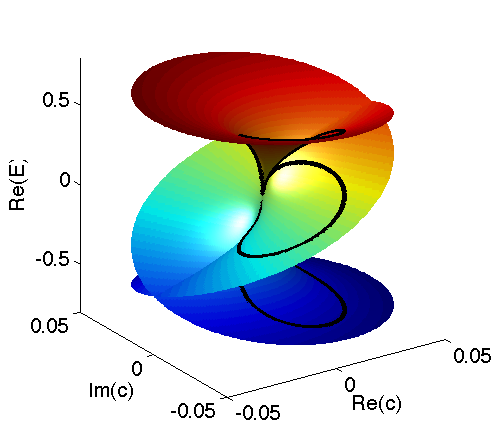}
\includegraphics[width=4cm]{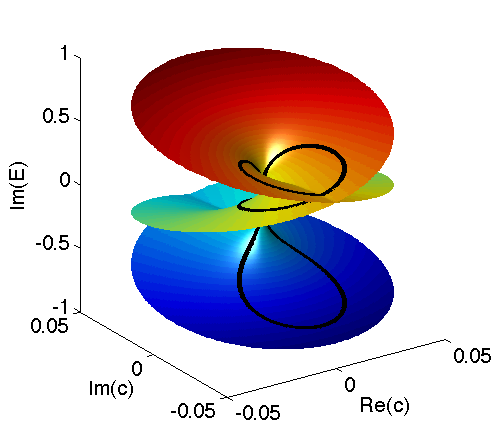}
\includegraphics[width=4cm]{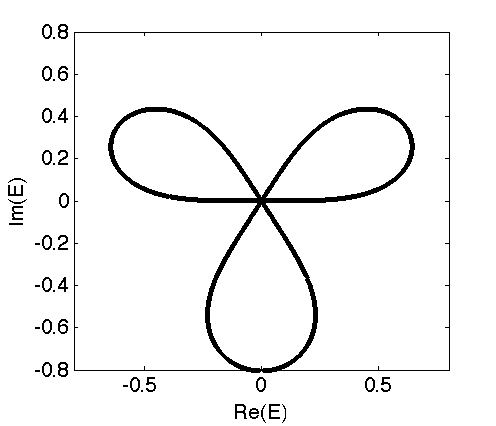}
\includegraphics[width=4cm]{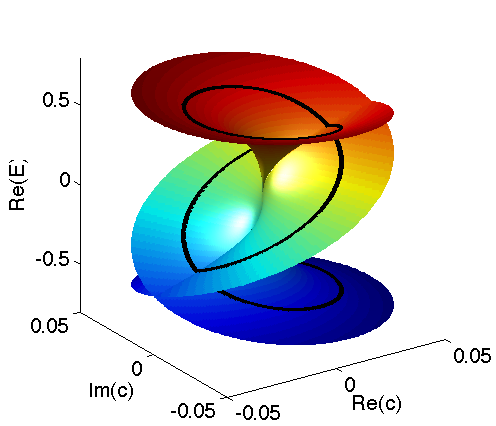}
\includegraphics[width=4cm]{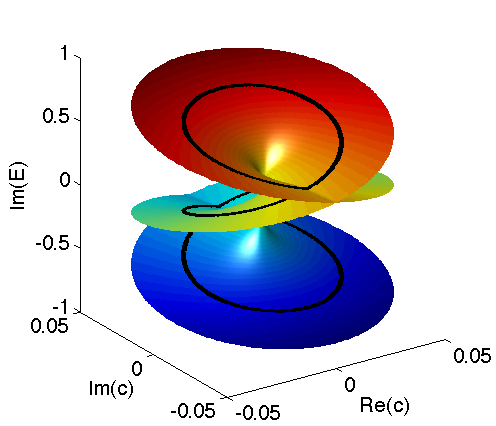}
\includegraphics[width=4cm]{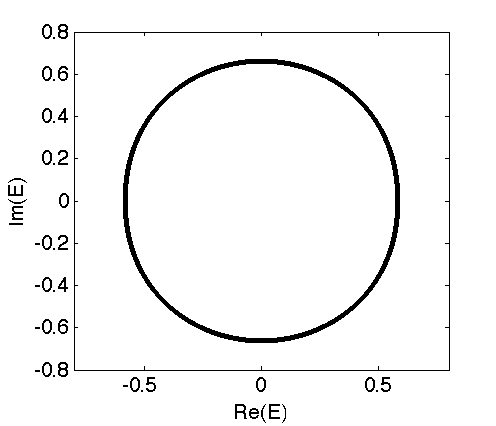}
\caption{Third-root behaviour of eigenvalues in dependence on a perturbation parameter around an EP3, and resulting eigenvalue patterns under a cyclic parameter variation. The left and the middle columns show the real and the imaginary parts, respectively, as functions of the complex perturbation parameter. The solid black line illustrates the eigenvalue curves that are observed when the parameter is varied along a one-dimensional closed loop. The right column shows the corresponding eigenvalue trajectories in the complex plane. Three distinct behaviours are observed for a loop outside the EP (top), passing through the EP (middle), and encircling the EP (bottom).}
\label{fig2}
\end{figure}
\begin{figure}[tb!]
\centering%
\includegraphics[width=4cm]{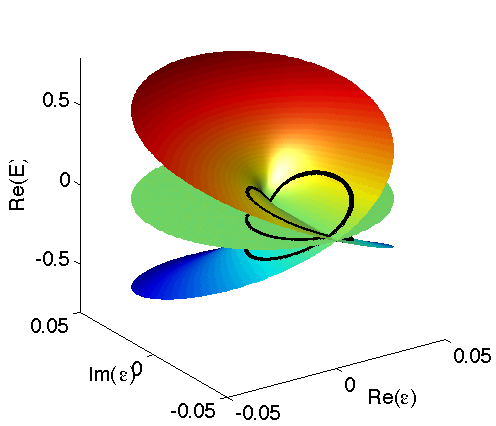}
\includegraphics[width=4cm]{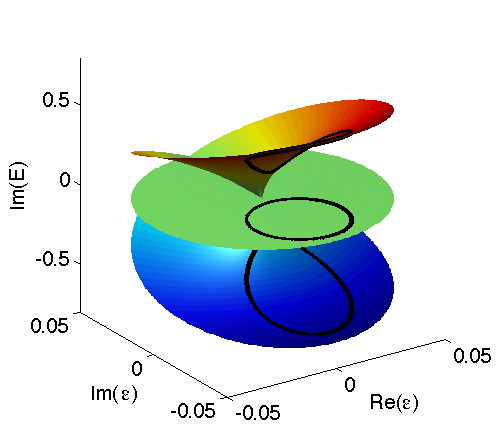}
\includegraphics[width=4cm]{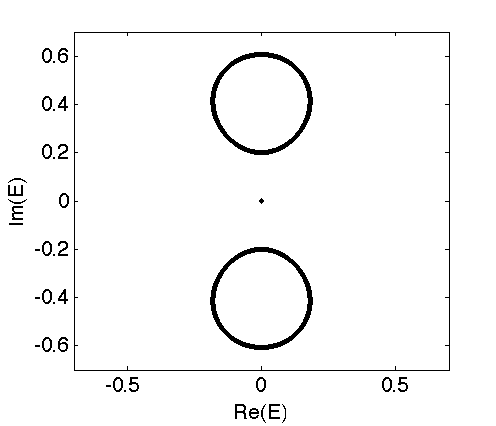}
\includegraphics[width=4cm]{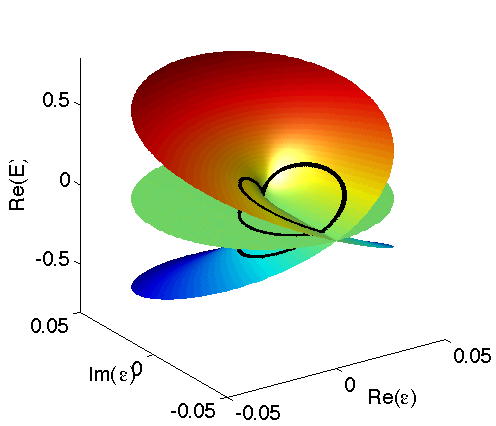}
\includegraphics[width=4cm]{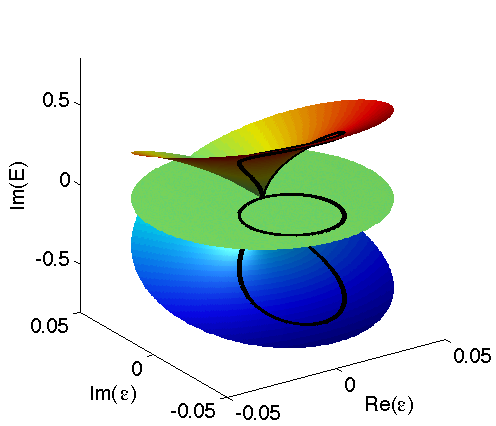}
\includegraphics[width=4cm]{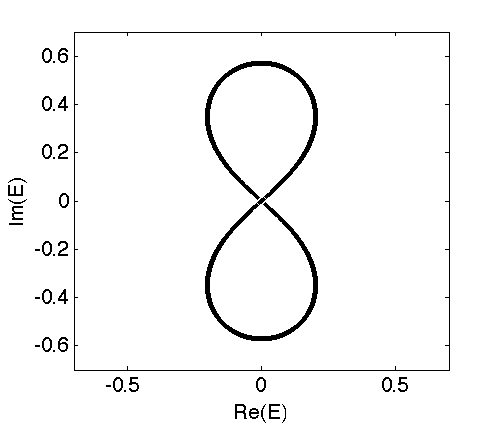}
\includegraphics[width=4cm]{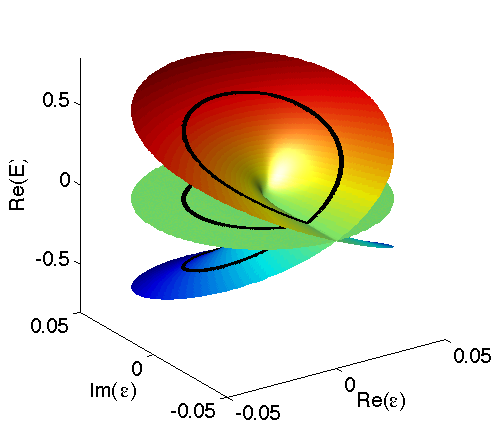}
\includegraphics[width=4cm]{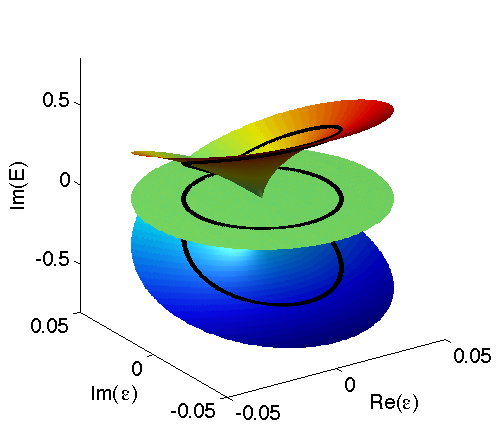}
\includegraphics[width=4cm]{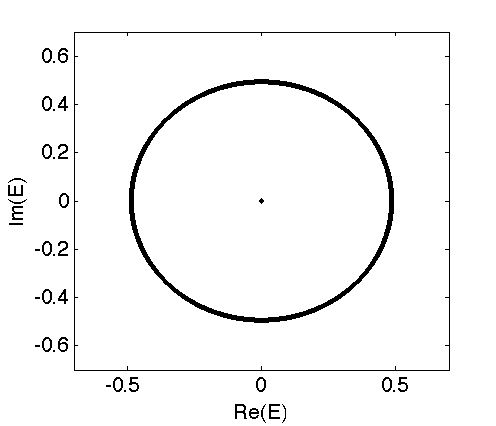}
\caption{As in figure \ref{fig2}, however, for a square-root dependence on the perturbation parameter.}
\label{fig3}
\end{figure}

One might argue that for an EP2 this is of little consequence, as the only two possibilities in this case are a square-root expansion and a conventional Taylor expansion, where the latter can be viewed as a special case of the former for which all odd coefficients vanish. For degeneracies of more than two eigenvalues and eigenfunctions, however, the situation is quite different. In particular, it follows that for the case of an EP3 there are two possible power expansions (apart from the Taylor expansion, which can be viewed as a special case of one of the other expansions) for the eigenvalues in the perturbation parameter $z$. The first (better known) expansion consists of a single cycle with the three perturbed eigenvalues being the three branches of a third root expansion. In the second (less appreciated) scenario one of the eigenvalues has a conventional Taylor expansion, and the remaining two are given by the branches of a square-root expansion. This second scenario is similar to the characteristic behaviour of an EP2. Typical examples of the two cases are depicted in figure \ref{fig1} in the top and bottom panel, respectively. The figure shows the real and the imaginary parts of the eigenvalues of a $3\times 3$ matrix perturbed around an EP3 as a function of the complex perturbation parameter. The situation in the upper panel corresponds to a third-root behaviour, while the situation in the lower panel corresponds to the second scenario. Note that in the latter case one of the eigenvalues is independent of the perturbation parameter up to the leading order of the other two eigenvalues.
 
This perturbative behaviour gives rise to typical patterns of eigenvalues for cyclic parameter variation. It is well known that for an EP2, due to the square-root behaviour, the two eigenvalues interchange when the EP is surrounded and form a single closed loop in the complex plane, while they describe two seperate closed curves if the EP is not enclosed. In the case of an EP3 the typical patterns of eigenvalues under parameter variation for the two possible scenarios are depicted in figures \ref{fig2} and \ref{fig3}, respectively. When encircling an EP3 along a parameter loop that corresponds to a triple-root perturbation, the three eigenvalues interchange, which results in a single closed curve formed by the three eigenvalues in the complex plane depicted in the lower right panel of figure \ref{fig2}. This is the standard behaviour usually connected to an EP3. However, the second perturbation scenario yields eigenvalue patterns commonly connected with EP2s, as depicted for an example in figure \ref{fig3}. 

The explicit perturbative expressions for the eigenvalues and eigenvectors can be obtained from a generalised perturbation method \cite{Seyr03}, by inserting an ansatz for the series expansion into the eigenvalue equation and equating powers of the perturbation paramter $z$. This yields the following leading order behaviour for an EP2: 
\begin{eqnarray}
\lambda&=\lambda_0+\lambda_{1}z^{1/2}+o(z^{1/2}),\quad \rm{with}\quad\lambda_{1}=(\vect{u_0}^T \hat H_1 \vect{u_0})^{1/2},\\
\vect{u}&=\vect{u_0}+\lambda_{1}\vect{u_1}z^{1/2}+o(z^{1/2}).
\end{eqnarray}
That is, the leading order correction for the eigenvectors is collinear to the first Jordan vector $\vect{u_1}$ at the EP. This is in fact similar for an EP3, where in the case of  a third root series expansion one finds:
\begin{eqnarray}
\lambda&=\lambda_0+\lambda_{1}z^{1/3}+o(z^{1/3}),\quad \rm{with}\quad\lambda_{1}=(\vect{u_0}^T \hat H_1 \vect{u_0})^{1/3},\\
\label{eqn_evc_EP3_trrt}\vect{u}&=\vect{u_0}+\lambda_{1}\vect{u_1}z^{1/3}+o(z^{1/3}).
\end{eqnarray}
For both cases the perturbation series has to be modified in the case $\vect{u_0}^T\hat H_1 \vect{u_0}=0$, which in fact corresponds to the condition $a_{n1}\neq 0$ in Theorem \ref{Th1}. For an EP2 the alternative behvaiour is given by a Taylor expansion in the perturbation parameter \cite{Seyr03}, which we shall not discuss here in further detail.

In the case of an EP3 the alternative behaviour is the square-root expansion of two of the eigenvalues and eigenvectors, and a Taylor expansion for the remaining eigenvalue and eigenvector, which explicitly yields:
\begin{eqnarray}
\lambda&=\left\{
\begin{array}{ll}
\lambda_0\pm\lambda^{(1)}_{1}z^{1/2}+o(z^{1/2}),\quad \rm{with}\quad\lambda^{(1)}_{1}=(2 \vect{u_1}^T \hat H_1 \vect{u_0})^{1/2}\\
\\ 
\lambda_0+\lambda^{(2)}_1z+o(z^{1/2}),\quad \rm{with}\quad\lambda^{(2)}_{1}=\frac{\vect{u_0}^T\hat H_1\hat G^{-1}\hat H_1 \vect{u_0}}{2\vect{u_0}^T\hat H_1\vect{u_1}}
\end{array}\right.\,\\
\nonumber\\
\label{eqn_evc_EP3_sqrt}\vect{u}&=\left\{
\begin{array}{ll}
\vect{u_0}+\lambda^{(1)}_{1}\vect{u_1}z^{1/2}+o(z)\\
\vect{u_0}+(\lambda^{(2)}_1\vect{u_1}-\hat G^{-1}\hat H_1 \vect{u_0})z+o(z)\, ,
\end{array}\right.
\end{eqnarray}
where the invertible operator $\hat G=\hat H_0-\lambda_0\mathds{1}-\vect{u_0} \vect{u_2}^T$ has been introduced by removing the kernel of the noninvertible operator $\hat H_0-\lambda_0\mathds{1}$.

The fact that the leading order perturbations of the eigenvectors are branches of an analytic function leads to an interchange of the eigenvectors under cyclic parameter variation. While the square-root expansion of the second EP3 scenario leads to an interchange between two eigenvalues and the two corresponding eigenvectors after one cycle just as in the case of an EP2, the phase that is acquired by the eigenvectors after two cycles, is different in these two cases, due to the different Jordan block structure at the EP. In what follows we shall discuss this behaviour in more detail.

\section{Behaviour of the eigenvectors under cyclic parameter variations}
\label{sec_geo}
In Hermitian quantum mechanics, a system initially in an eigenstate returns to this initial state after an adiabatic cyclic variation of parameters around an eigenvalue degeneracy. During this cyclic variation it acquires an additional geometric phase; the celebrated Berry-phase \cite{Berr84b}. In the presence of EPs, however, the situation is more intricate. It is a well known result, which we shall briefly review in what follows, that surrounding an EP2 in parameter space, the two eigenvectors interchange after one cycle and two cycles are needed to return to the initial eigenstate. In addition the states pick up a phase of $\pi$ after two cycles \cite{Heis99,Heis01,03crossing,Mail05,Mehr08,Demb01}. Another important difference to the Hermitian case has been pointed out recently in \cite{Uzdi11,Berr11,Berr11b}: In the presence of non-Hermiticities the adiabatic theorem might break down in the sense that non-adiabatic corrections do not have to vanish. As a consequence fundamental difficulties for the observation of the phases or even the interchange behaviour using adiabatic parameter variation are to be expected. Here we shall for the moment not be concerned with this issue, and first analyse the situation for abstract cyclic parameter variations.

In this section we shall investigate the effect of a cyclic variation of parameters around an EP3. In particular, we consider the variation of one complex parameter around a loop in the complex plane encircling the EP3. We find that the geometric phase is determined only by the order of the EP, while the number of parameter cycles that are necessary for the eigenvectors to return to their initial state is governed by the Puiseux expansion form of the eigenvectors, and can thus depend on the perturbation type around an EP3. The situation for more general parameter perturbations is briefly discussed in Section \ref{sec_gen}.

To begin with, let us review the situation for EP2s. 
The leading order of the eigenvectors in the neighborhood of the EP in dependence on the complex parameter $z$, which we shall decompose into its amplitude and phase, $z=r\rme^{\rmi \phi}$, has the form
\begin{equation}
\vect{u_{\pm}}=\vect{u_0} \pm \sqrt{r}\alpha e^{\rmi \phi/2}\vect{u_1},
\end{equation}
where $\alpha\in\mathds{C}$ is constant. 

To determine the behaviour under a cyclic variation of the parameter $z$ it is necessary to fix the normalisation of the eigenvectors consistently for all parameters. In particular to obtain meaningful results for the phase, care has to be taken that the parallel transport condition (that the derivative of the eigenfunction with respect to the parameter be orthogonal to the eigenfunction itself) is fulfilled. Here we use the standard normalisation for the eigenvectors of non-Hermitian operators: $\vect{u}_{\pm}^T\vect{u_{\pm}}=1$, which also automatically ensures the parallel transport condition.
Using the self-orthogonality condition (\ref{eqn_J1_EP2}) and the Jordan chain relations (\ref{eqn_J2_EP2}) we find in leading order of $r$ that
\begin{equation}
\vect{u_{\pm}}^T\vect{u_{\pm}}=\pm 2 \alpha \sqrt{r}e^{\rmi\phi/2},
\end{equation}
and thus we get a suitable expression for $\vect{u_\pm}$:
\begin{equation}
\label{eqn_EP2_eigvec_norm}
\vect{u_{\pm}}=\frac{\vect{u_0}\pm \alpha r^{1/2}e^{\rmi\phi/2}\vect{u_1}}{\sqrt{\pm2\alpha \sqrt{r}e^{\rmi\phi/2}}}.
\end{equation}
When the parameter performs one cycle, that is, when the angle $\phi$ traverses the interval $\left[ 0, 2\pi\right]\,$, the numerators in (\ref{eqn_EP2_eigvec_norm}) interchange. This phenomenon is purely due to the square root perturbation behaviour of the eigenvectors. From this it follows that up to a phase, which is determined by the denominators, the eigenvectors return to their initial state after two cycles of parameters in the case of an EP2. We will see shortly that a similar behaviour can prevail for an EP3. Now we turn to the behaviour of the denominator in (\ref{eqn_EP2_eigvec_norm}). Recalling that the complex square root displays a half line branch cut in the complex plane, one finds that either the positive or the negative argument in the square root passes through the branch cut. This induces a minus sign to one of the denominators when going through one cycle, and a minus sign to the other when passing through the second cycle. This corresponds to the geometric phase of $\pi$ that the eigenvectors acquire after two cycles. In summary, combining the behaviours of both the denominator and the numerator we obtain the well-known result for the behaviour of the eigenvectors under successive cyclic parameter variations (denoted by $\circlearrowright$):
\begin{equation}
\vect{u}_{+}\circlearrowright \vect{u}_{-} \circlearrowright -\vect{u}_{+}\circlearrowright-\vect{u}_{-}\circlearrowright \vect{u}_{+}.
\end{equation}
The two eigenvectors interchange for each cycle, where one picks up an additional minus sign. Thus, after two cycles the system returns to its initial eigenvector up to a geometric phase of $\pi$. This phase is intrinsically related to the Jordan chain structure at the exceptional point.

We now proceed in an analogous manner to determine the behaviour in the presence of EP3s. We begin with the standard triple root case (\ref{eqn_evc_EP3_trrt}), where the three relevant eigenvectors in leading order in the parameter $z=r\rme^{\rmi \phi}$ have the form 
\begin{equation}
\vect{u^{(k)}}=\vect{u_0} +\alpha\rme^{\rmi k2\pi/3}{r}^{1/3}\, \rme^{\rmi \phi/3}\vect{u_1}. 
\end{equation}
The superscript $\bf(k)$ denotes the different branches of the complex third root. Choosing again the standard normalisation $(\vect{u^{(k)}})^T\vect{u^{(k)}}=1$, and using the Jordan chain properties (\ref{eqn_J1_EP3}) and (\ref{eqn_J2_EP3}), we find the following expression for the normalised eigenvectors in leading order of $r$:
\begin{equation}
\label{eqn-eigenvec-EP3-1}
\vect{u^{(k)}}=\frac{\vect{u_0}+\alpha\rme^{\rmi k2\pi/3}{r}^{1/3}\, \rme^{\rmi \phi/3}\vect{u}_1}{\sqrt{\alpha\rme^{\rmi k4\pi/3}r^{2/3}\, \rme^{2\rmi\phi/3}}}.
\end{equation}
We observe that the numerators are cyclically interchanged in each cycle. Thus, after three cycles, the eigenvectors return to themselves, up to a possible phase, which is determined by the behaviour of the denominator in (\ref{eqn-eigenvec-EP3-1}). The crucial part in the denominator is the factor $\sqrt{\rme^{\rmi k4\pi/3}\, \rme^{2\rmi \phi/3}}$. When $\phi$ describes a full cycle, each of the arguments in the square root describes two thirds of a full circle in the complex plane. As a 
consequence, in each cycle two of the arguments pass through the branch cut,
and thus, the denominators cyclically interchange and two of the three pick up a minus sign. If we choose the signs of the eigenvectors in the initial configuration appropriately, this minus sign can be absorbed in the definition of the eigenvectors, and we find the following pattern for successive parameter cycles:
\begin{equation}
\vect{u^{(1)}}\circlearrowright \vect{u^{(2)}} \circlearrowright \vect{u^{(3)}}\circlearrowright \vect{u^{(1)}}.
\end{equation}
The three eigenvectors cyclically interchange for each parameter cycle and thus return to their initial state after three cycles, where none of them picks up a geometric phase. While the three cycles required to recover the original configuration are related to the triple root structure of the parameter dependence, the fact that no phase is acquired is due to the Jordan block structure at the EP3, and we will find the same behaviour for the second perturbation scenario (\ref{eqn_evc_EP3_sqrt}). In this case we first consider the two eigenvectors with the square root behaviour, which take the same form as in the EP2 case:
\begin{equation}
\vect{u_{\pm}}=\vect{u_0} \pm \alpha\sqrt{r} \rme^{\rmi \phi/2}\vect{u_1}.
\end{equation}
However, due to the different Jordan chain relations at the EP3, instead of (\ref{eqn_EP2_eigvec_norm}) we find for the normalised eigenvectors in the lowest order of $r$:
\begin{equation}
\label{eqn_eigvec_EP3_sqrt_norm}
\vect{u}_{\pm}=\frac{\vect{u_0}\pm\rme^{\rmi \phi/2}\vect{u}_1}{\sqrt{\alpha^2 r\, \rme^{\rmi \phi}}}.
\end{equation}
As in the case of the EP2, the numerators interchange in each parameter cycle. 
The behaviour of the normalisation, however, differs from the EP2 case. For each cycle of paramaters the argument in the square root of the denominator in (\ref{eqn_eigvec_EP3_sqrt_norm}) performs a whole cycle in the complex plane, and hence the denominator picks up a minus sign. 
Again we can absorb the sign by an appropriate choice of the initial eigenvectors. Putting these together, we find for the behaviour of two of the eigenvectors under consecutive parameter cycles:
\begin{equation}
\vect{u}_{+}\circlearrowright \vect{u}_{-} \circlearrowright \vect{u}_{+}
\end{equation}
The two eigenvectors interchange for each cycle of parameters and thus return to their initial state after two cycles just as in the EP2 case, since this interchange is encoded in the Puiseux expansion. However, in contrast to the EP2 case, none of the eigenvectors picks up a geometric phase. A study of the remaining eigenvector in a similar way shows, that it returns to itself after each cycle while picking up a minus sign, that is, a phase of $\pi$. 

In summary, we found that the form of the Puiseux expansion of the eigenvectors determines the number of parameter cycles that are necessary for the eigenvectors to return to their initial state, whereas the order $N$ of the degeneracy determines the geometric phase. Note that the results here are based on the perturbation around the degeneracy of three eigenfunctions, which is not restricted by the actual dimensionality of the considered Hilbert space. However, modifications for larger circles in parameter space, for which higher orders of the perturbation expansion become relevant are possible \cite{Mail05}. Finally, the observed phenomena should directly generalise to degeneracies of more than three eigenvectors.

\section{A physical example: EP3s in three coupled wave guides}
\label{sec_ex}
Let us now present an experimental setup in the context of optics, for which the two different behaviours of an EP3 might be observable. In particular, we consider light propagation in three approximately parallel coupled wave guides. In the paraxial approximation the spatial light propagation is described by an equation analogous to the time dependent Schr\"odinger equation, where one spatial direction takes the role of the time, and the potential is replaced by the refractive index of the wave guides, which can be complex in the case of absorbing or amplifying structures. Recently an EP2 has been experimentally observed in a similar structure of two wave guides \cite{Guo09,Ruet10}. For more details on the analogy between quantum dynamics and optical structures, see, e.g., \cite{Long08,Long11,Makr08,Klai08}.

The three wave guide system we consider here is constructed to resemble a quantum system with a Hamiltonian of the form 
\begin{equation}
\label{eqn_example_Ham}
\hat H=\left(\begin{array}{ccc} 
a-2\rmi\gamma & \sqrt{2}\, v & 0\\
\sqrt{2}\, v & 0 & \sqrt{2}\, v \\
0 & \sqrt{2}\, v & b+2\rmi\gamma
\end{array}\right),
\end{equation}
with $\gamma,v\in\mathds{R}$, and $a,b\in\mathds{C}$. This corresponds to a situation where each of the wave guides supports a single mode. The real and the imaginary parts of the diagonal elements are related to the refractive index and the absorption and amplification in the wave guides, and can thus be changed by choosing different materials, or varying the amplification strength or depths of the wave guides (within a range for which the single mode approximation remains valid). Neighbouring wave guides are coupled via the evanescent field between them. Thus, the coupling parameter $v$ is related to the distance of the wave guides (the coupling is large if the wave guides are close to each other and approaches zero, if they are far apart). The numerical values appearing in the Hamiltonian have been chosen for convenience such that the system displays an EP3 for the values $\gamma=v$ and $a=0=b$. Since the time dependence in the Schr\"odinger equation corresponds to a spatial variation in the optical context, an adiabatic variation of parameters can be achieved by a variation of the wave guide parameters along the beam propagation direction on a length scale that is long compared to the intrinsic scales of the system. The parameters necessary for an observation of the predicted phenomena should be within experimental reach, since the loop surrounding the exceptional point, and thus the corresponding parameter variations, can be arbitrarily small. Furthermore, the presence of the EP3 relies only on the ratio of absorption and coupling, and the differences in the refractive indices of the three waveguides, but not on absolute values. In the context of Hermitian quantum dynamics many comparative models with adiabatic time dependence have been realised in wave guide structures \cite{Long08}. However, it has recently been shown \cite{Uzdi11,Berr11,Berr11b} that the non-Hermiticity, that is, the presence of loss and gain in the wave guides, can destroy the expected adiabaticity. This might prevent a measurement of the interchange behaviour and the phases under adiabatic parameter variation. Instead one might have to be contended with the measurement of instantaneous eigenvalues and modes. 

The two different scenarios discussed in the present paper can in principle be observed in a system described by the Hamiltonian (\ref{eqn_example_Ham}). It can easily be verified that the eigenvalues and eigenvectors of (\ref{eqn_example_Ham}) around the EP3 have the typical third root dependence on the complex parameter $z$ if we choose $a=b=z$, whereas they show the square root behaviour for the choice $a=-b=z$. In fact the two different cases depicted in figure \ref{fig1} are the eigenvalues of the present example, where $\gamma=v=1$. 

\section{Outlook: More general two-parameter perturbations}\label{sec_gen}
The results presented here rely on the Puiseux expansion in the neighbourhood of degeneracies. It is interesting to note that in contrast to Taylor expansions around nondegenerate eigenvalues, there are mathematical subtleties leading to qualitatively new behaviour for multi-paramter perturbations around Jordan blocks. This is connected to the fact that there is no direct generalisation of Puiseux expansions to several variables (see, e.g., \cite{Aroc10}). Thus, if parameters are varied in a way which cannot be combined into a single complex parameter, the behaviour cannot be captured by the perturbative considerations presented above. A detailed analysis of the possible structures for these cases, however, goes beyond the scope of the present paper, and is an interesting topic for future investigations. Here, we shall only present an example to illustrate some structures that might be expected.  

\begin{figure}[tb!]
\centering%
\includegraphics[width=4cm]{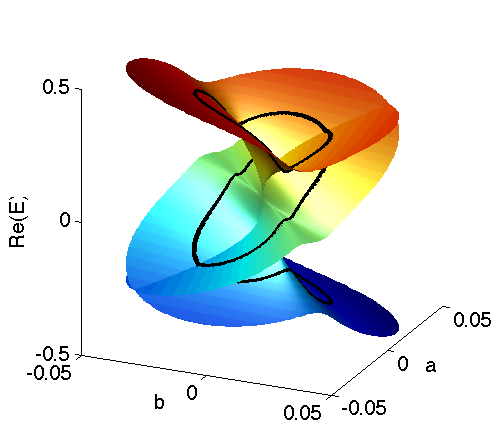}
\includegraphics[width=4cm]{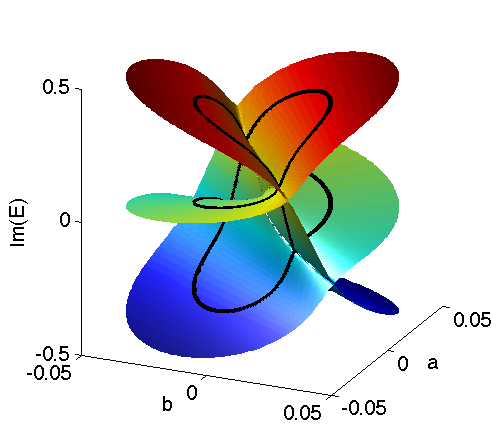}
\includegraphics[width=4cm]{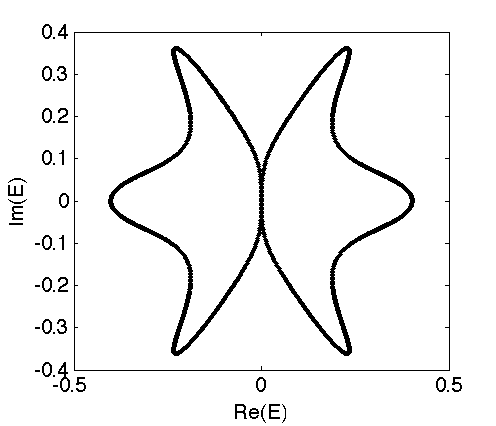}
\caption{Real (left) and imaginary (right) parts of the eigenvalues of the Hamiltonian (\ref{eqn_example_Ham}) for $\gamma=1=v$ in dependence on the parameters $a$ and $b$. The solid line shows the eigenvalues along a curve in parameter space that encircles the exceptional point. The figure on the right shows the corresponding eigenvalue trajectory.}
\label{fig4}
\end{figure}

For this purpose we return to the model (\ref{eqn_example_Ham}), and consider a variation of the independent variables $a,b\in\mathds{R}$, which cannot be combined into a single effective complex parameter. In figure \ref{fig4} we show the resulting eigenvalues for $\gamma=1=v$. It can be seen that although the eigenvalues form a three-sheeted Riemann surface, we do not find the typical structures discussed above of either a triple root or a square root and an additional constant eigenvalue. Instead, the eigenvalues form a more intricate structure with multiple additional connections, which resembles a square-root behaviour in dependence on the parameter $b$ in the plane $a=0$ and a triple-root behaviour in dependence on the parameter $a$ in the plane $b=0$. The figure also shows the resulting eigenvalue pattern for a cyclic parameter variation $(a,b)=r(\cos(\phi),\sin(\phi))$, with a small constant $r\in\mathds{R}$ and $\phi\in[0,2\pi]$. Although the EP is surrounded, the resulting eigenvalue trajectories show a self-crossing. This is usually related to parameter loops crossing an EP, whereas here it results from the additional connections in the imaginary parts of the eigenvalues. The related behaviour of the eigenvectors is correspondingly more elaborate and will be discussed in a future study. 

\section{Discussion}
\label{sec_dis}
In summary, we have presented the behaviour of eigenvalues and eigenvectors of symmetric non-Hermitian operators in dependence on a complex parameter in the neighbourhood of exceptional points at which three eigenvalues and the corresponding eigenfunctions coalesce. In particular, we have studied the patterns resulting from cyclic parameter variations. We have shown that there are two basic scenarios, related to the two possible Puiseux expansions of the eigenvalues and eigenvectors. They lead to characteristic interchanges of eigenvalues and eigenfunctions for consecutive cycles. In addition, there can be geometric phases, which are related to the degeneracy itself and independent of the particular parameter variation. We have presented a possible experimental setup in which both behaviours might be observable. Finally, we have illustrated that the patterns resulting from general cyclic variation of parameters (which cannot be expressed as the variation of one effective complex parameter) can show additional qualitatively new features. Further interesting generalisations are expected for non-symmetric operators. An extension of the present study to exceptional points of higher orders, along the lines presented here, is a promising topic for future investigations.

\vskip 9pt
The authors would like to thank Dorje Brody and Raam Uzdin for valuable comments on the manuscript. EMG is grateful to Barbara Dietz for stimulating and useful discussions.   
EMG acknowledges support via the Imperial College Junior Research Fellowship scheme.


\vskip 10pt 

\begin{thebibliography}{999}
\bibitem{Seyr03}
Seyranian A~P and Mailybaev A~A 2003 {\em Multiparameter stability theory with
  mechanical applications\/} (Singapore: World Scientific)

\bibitem{Kiri07}
Kirillov O~N 2007 {\em Int. J. Nonlinear Mech.\/} {\bf 42} 71

\bibitem{Panc55}
Pancharatnam S 1955 {\em Proc. Indian Acad. Sci., Sect. A\/} {\bf 42} 86

\bibitem{Berr98b}
Berry M~V and O'Dell D~H~J 1998 {\em J. Phys. A\/} {\bf 31} 2093

\bibitem{Berr03}
Berry M~V and Dennis M~R 2003 {\em Proc. Roy. Soc. Lond. A\/} {\bf 459} 1261

\bibitem{Klai08}
Klaiman S, G\"{u}nther U and Moiseyev N 2008 {\em Phys. Rev. Lett.\/} {\bf 101}
  080402

\bibitem{Wier08}
Wiersig J, Kim S~W and Hentschel M 2008 {\em Phys. Rev. A\/} {\bf 78} 053809

\bibitem{Dett09}
Dettmann C~P, Morozov G~V, Sieber M and Waalkens H 2009 {\em Phys. Rev. A\/}
  {\bf 80} 063813

\bibitem{Lee09}
Lee S~B, Yang J, Moon S, Lee S~Y, Shim J~B, Kim S~W, Lee J~H and An K 2009 {\em
  Phys. Rev. Lett.\/} {\bf 103} 134101

\bibitem{Long10}
Longhi S 2010 {\em Phys. Rev. A\/} {\bf 81} 022102

\bibitem{Grae11}
Graefe E~M and Jones H~F 2011 {\em Phys. Rev. A\/} {\bf 84} 013818

\bibitem{Mois11book}
Moiseyev N 2011 {\em Non-{Hermitian} {Quantum} {Mechanics}\/} (Cambridge:
  Cambridge University Press)

\bibitem{Berr04}
Berry M~V 2004 {\em Czech. J. Phys.\/} {\bf 54} 1039

\bibitem{Bend99a}
Bender C~M, Boettcher S and Meisinger P~N 1999 {\em J. Math. Phys.\/} {\bf 40}
  2201

\bibitem{Bend02b}
Bender C~M, Brody D~C and Jones H~J 2002 {\em Phys. Rev. Lett.\/} {\bf 89}
  270401

\bibitem{Demb01}
Dembowski C, {Gr\"af} H~D, Harney H~L, Heine A, Heiss W~D, Rehfeld H and
  Richter A 2001 {\em Phys. Rev. Lett.\/} {\bf 86} 787

\bibitem{Demb04}
Dembowski C, Dietz B, {Gr\"af} H~D, Harney H~L, Heine A, Heiss W~D and Richter
  A 2004 {\em Phys. Rev. E\/} {\bf 69} 056216

\bibitem{Diet07}
Dietz B, Friedrich T, Metz J, Miski-Oglu M, Richter A, {Sch\"afer} F and
  Strafford C 2007 {\em Phys. Rev. E\/} {\bf 75} 027201

\bibitem{Diet11}
Dietz B, Harney H~L, Kirillov O~N, Miski-Oglu M, Richter A and Sch\"afer F 2011
  {\em Phys. Rev. Lett.\/} {\bf 106} 150403

\bibitem{Guo09}
Guo A, Salamo G~J, Duchesne D, Morandoti R, Volatier-Ravat M, Aimez V,
  Siviloglou G~A and Christodoulides D~N 2009 {\em Phys. Rev. Lett.\/} {\bf
  103} 093902

\bibitem{Ruet10}
R{\"u}ter C~E, Makris K~G, El-Ganainy R, Christodoulides D~N, Segev M and Kip D
  2010 {\em Nature Physics\/} {\bf 6} 192

\bibitem{Kato66book}
Kato T 1966 {\em Perturbation theory for linear operators\/} (Berlin: Springer
  Verlag)

\bibitem{Bend69}
Bender C~M and Wu T~T 1969 {\em Phys. Rev.\/} {\bf 184} 1231

\bibitem{Mois78}
Moiseyev N, Certain P R and Weinhold F 1978 {\em Molec. Phys.\/} {\bf 36}
  1613

\bibitem{Heis99}
Heiss W~D 1999 {\em Eur. Phys. J. D\/} {\bf 17} 1

\bibitem{Heis01}
Heiss W~D and Harney H~L 2001 {\em Eur. Phys. J. D\/} {\bf 17} 149

\bibitem{03crossing}
Keck F, Korsch H~J and Mossmann S 2003 {\em J. Phys. A\/} {\bf 36} 2125

\bibitem{Mail05}
Mailybaev A~A, Kirillov O~N and Seyranian A~P 2005 {\em Phys. Rev. A\/} {\bf
  72} 014104

\bibitem{Mehr08}
Mehri-Dehnavi H and Mostafazadeh A 2008 {\em J. Math. Phys.\/} {\bf 49} 082105

\bibitem{Cart07}
Cartarius H, Main J and Wunner G 2007 {\em Phys. Rev. Lett.\/} {\bf 99} 173003

\bibitem{Heis10}
Heiss W~D 2010 {\em Eur. Phys. J. D\/} {\bf 60} 257

\bibitem{Zhen10}
Zheng M~C, Christodoulides D~N, Fleischmann R and Kottos T 2010 {\em Phys. Rev.
  A\/} {\bf 82} 010103
  
\bibitem{Cart11}
Cartarius H and Moiseyev N 2011 {\em Phys. Rev. A\/} {\bf 84} 013419

\bibitem{Atab11} 
Atabek O, Lefebvre R, Lepers M, Jaouadi A, Dulieu O and Kokoouline V 2011{\em Phys. Rev. Lett.} {\bf 106} 173002

\bibitem{Else11}
Elsen C, Rapedius K, Witthaut D and Korsch H~J 2011 {\em J. Phys. B} {\bf 44} 225301

\bibitem{Andr07}
Andrianov A~A, Cannata F and Sokolov A~V 2007 {\em Nucl. Phys. B} {\bf 773} 107

\bibitem{Cart09}
Cartarius H, Main J and Wunner G 2009 {\em Phys. Rev. A} {\bf 79} 053408

\bibitem{Heis08}
Heiss W~D 2008 {\em J. Phys. A\/} {\bf 41} 244010

\bibitem{08PT}
Graefe E~M, G\"unther U, Korsch H~J and Niederle A~E 2008 {\em J. Phys. A\/}
  {\bf 41} 255206

\bibitem{Ryu11}
Ryu J~W, Lee S~Y and Kim S~W 2011 {\em \phantom{0}\/} {\bf \phantom{0}}
  arXiv:1109.4216

\bibitem{Moro97}
Moro J, Burke J~V and Overton M~L 1997 {\em SIAM J. Matrix Anal. Appl.\/} {\bf
  18} 793

\bibitem{Ma98}
Ma Y and Edelman A 1998 {\em Linear Algebra Appl.\/} {\bf 273} 45

\bibitem{Uzdi11}
Uzdin R, Mailybaev A and Moiseyev N 2011 {\em J. Phys. A\/} {\bf 44} 435302

\bibitem{Berr11}
Berry M~V and Uzdin R 2011 {\em J. Phys. A\/} {\bf 44} 435303

\bibitem{Berr11b}
Berry M~V 2011 {\em J. Opt.} {\bf 13} 115701

\bibitem{Berr84b}
Berry M~V 1984 {\em Proc. R. Soc. Lond A\/} {\bf 392} 45

\bibitem{Long08}
Longhi S 2008 {\em Laser and Photon. Rev.\/} {\bf 3} 243

\bibitem{Long11}
Longhi S 2011 {\em J. Phys. A\/} {\bf 44} 051001

\bibitem{Makr08}
Makris K~G, El-Ganainy R, Christodoulides D~N and Musslimani Z~H 2008 {\em
  Phys. Rev. Lett.\/} {\bf 100} 103904

\bibitem{Aroc10}
Aroca F, Ilardi G and de~Medrano L~L 2010 {\em Int. J. Math.\/} {\bf 21} 1439

\end{thebibliography}
\end{document}